\documentclass{camera}
\usepackage{graphicx}  

\def\EPJ{Eur. Phys. J. }

\def\NP{Nucl. Phys. }
\def\PL{Phys. Lett. }

\begin{document}


\renewcommand{\thefootnote}{\fnsymbol{footnote}}

\title{
TWO-PHOTON PHYSICS AT LEP
\footnote{Presented at IFAE 2003, Lecce, Italy, 23-26 April, 2003.}
}

%
\author{Saverio Braccini}

%
\organization{
Istituto Nazionale di Fisica Nucleare,\\
Laboratori Nazionali di Frascati 
\footnote{now with TERA, Foundation for Onclogical Hadrontherapy,
CERN EP DIVISION, CH-1211 Geneva 23, Switzerland.}\\
E-mail: Saverio.Braccini@cern.ch
}

\maketitle

\abstract{
A remarkable number of studies have been performed at LEP
in the field of two-photon physics in the last four years. These results
represent a very important contribution to the understanding of strong interactions
at low energies. In particular, significant deviations from QCD predictions are
found in the cross sections of inclusive single particle, jet and beauty production.
A concise review of some of these results is presented.
}

%

\section{Introduction}

 Despite the fact that the LEP collider and the four detectors ALEPH, DELPHI, OPAL and L3
were  not conceived and constructed to study
two-photon interactions, a very remarkable number of results have been recently obtained
by the four LEP collaborations, leading to about twenty-five publications on this subject
in the last four years~\cite{Photon2003}.
It is important to notice that, at the high energies above the Z pole, photon-photon
interactions are characterized by a cross section more than two orders of
magnitude larger than the ${\rm e}^+{\rm e}^-$ annihilation process. Moreover,
a clean separation between the two-photon and
annihilation events is possible by using a cut in the visible energy. High energy
LEP data are therefore very well suitable for the study of two-photon interactions.

\begin{figure}[t]
\begin{center}
\includegraphics[width=.35\textwidth]{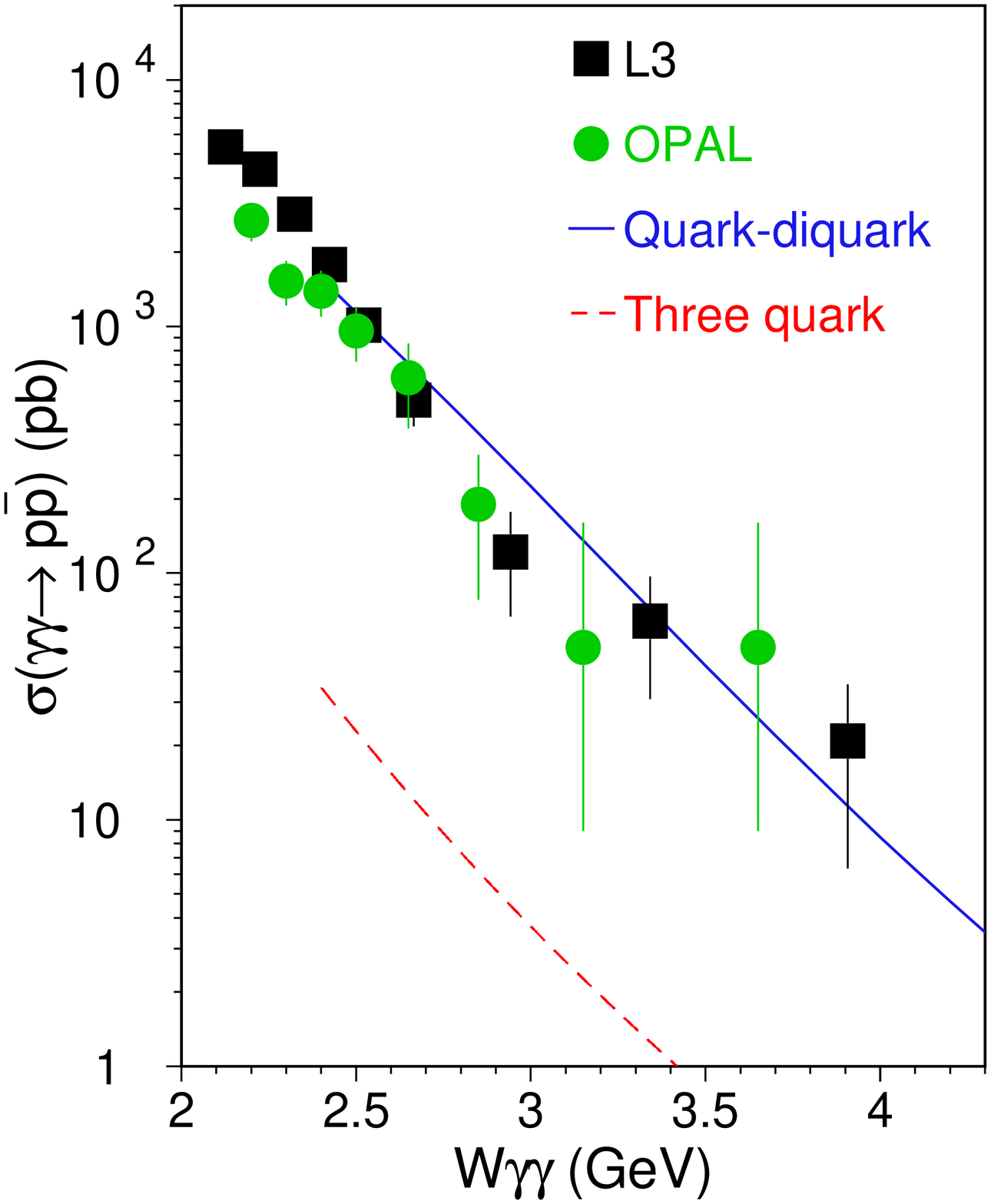}
\includegraphics[width=.50\textwidth]{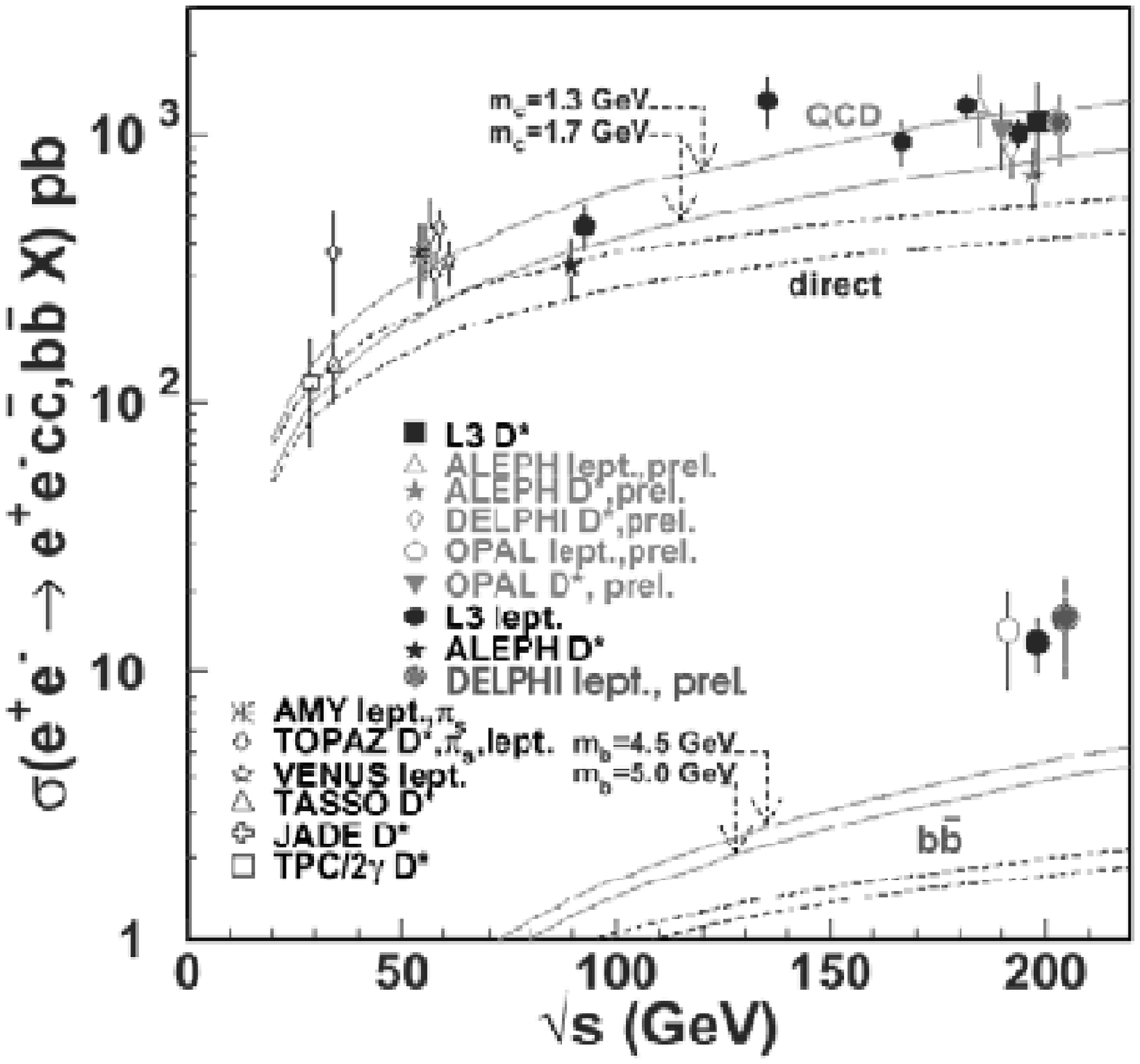}
\end{center}
\caption{The cross section as a function of the two-photon mass 
${\rm W}_{\gamma\gamma}$ for the process $\gamma\gamma\rightarrow
{\rm p}\bar{{\rm p}}$ (left). The cross section for open charm and beauty
production as a function of the ${\rm e}^+{\rm e}^-$ mass (right).}
\label{fig01} 
\end{figure}

\section{Exclusive processes}

 The study of the formation of light resonant states gives fundamental
information to light meson spectroscopy and glueball searches. Several
resonant states have been studied at LEP and their two-photon widths
have been measured with very high accuracy~\cite{Saverio-Meson2000}.
In particular, the formation of the f$_{\rm J}$(1710) has been observed for the first
time in gamma-gamma collisions by L3~\cite{L3-k0k0} and the spin-two tensor wave has been found to 
be dominant in this mass region. No signal has been observed for the $\xi$(2230) tensor
glueball candidate.

 Due to the high mass of charm quarks, the
formation of charmonia can be used to test perturbative
QCD predictions. The two-photon width of the $\eta_c$ and of the $\chi_{c2}$
has been measured by DELPHI~\cite{DELPHI-etac}, L3~\cite{L3-etac}~\cite{L3-chic} and OPAL~\cite{OPAL-chic}
and an upper limit for the formation of the $\eta_c'$ has been set by DELPHI~\cite{DELPHI-etacp}.

 The $\eta_b$ meson has not been observed so far and its search is very challenging.
Considering the very small number of expected events, only a combination of the data from
the four LEP experiments would allow the claim of a discovery. Up to present,
only one candidate has been observed ALEPH~\cite{ALEPH-etab} and preliminary results
have been presented by L3~\cite{Photon2003}.

 Baryon-antibaryon pair production allows to investigate the structure
of baryons which can be described in terms of three-quark or quark-diquark models.
The studies of  $\Lambda\bar{\Lambda}$ and $\Sigma^0\bar{\Sigma^0}$ by L3~\cite{L3-llbar}
and of ${\rm p}\bar{{\rm p}}$ exclusive production by L3~\cite{L3-pp}
and OPAL~\cite{OPAL-pp} clearly show a good agreement with the predictions of the 
quark-diquark model and exclude the three-quark model, as shown in 
Figure~\ref{fig01}~(left).

\begin{figure}[t]
\begin{center}
\includegraphics[width=.40\textwidth]{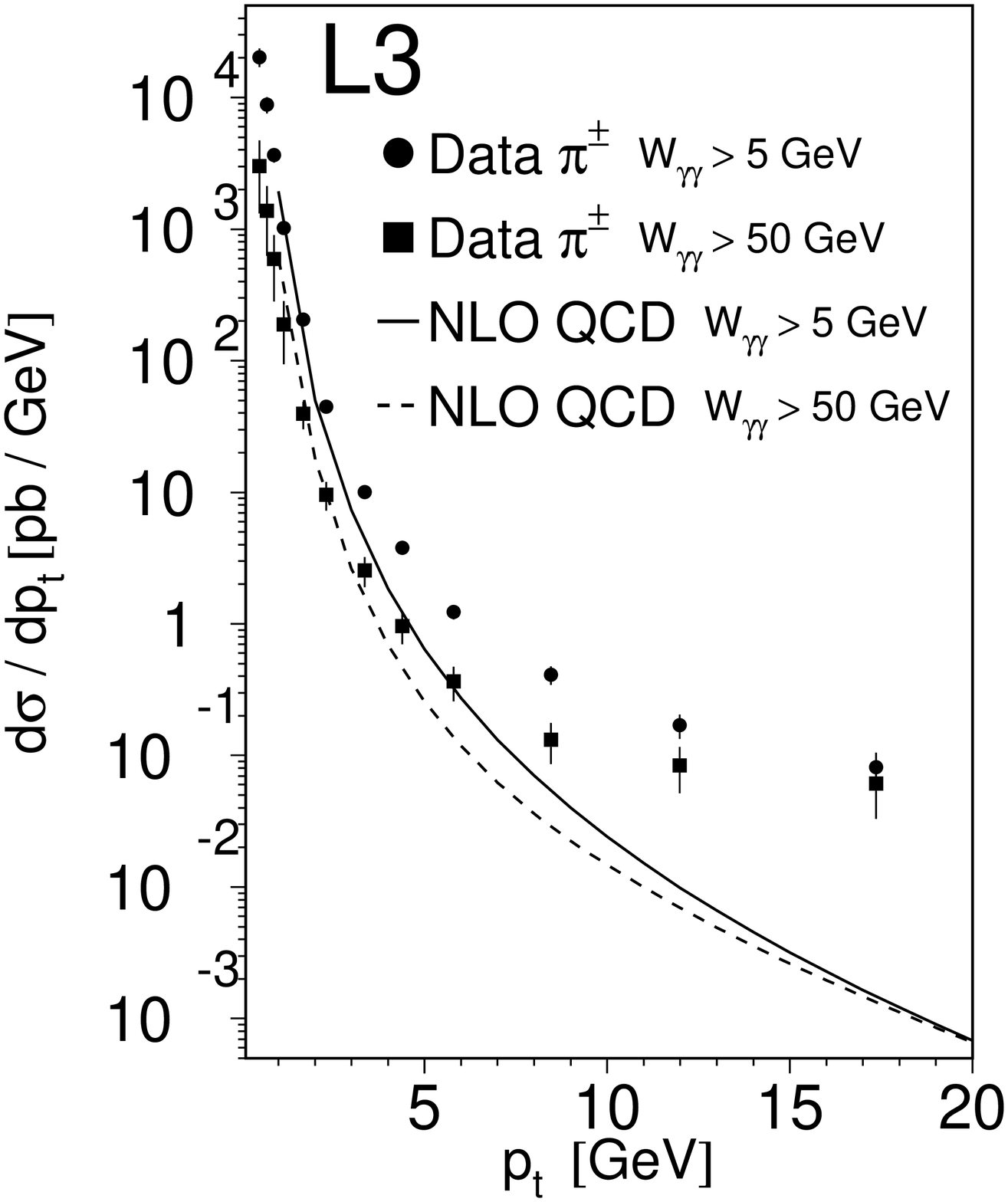}
\includegraphics[width=.40\textwidth]{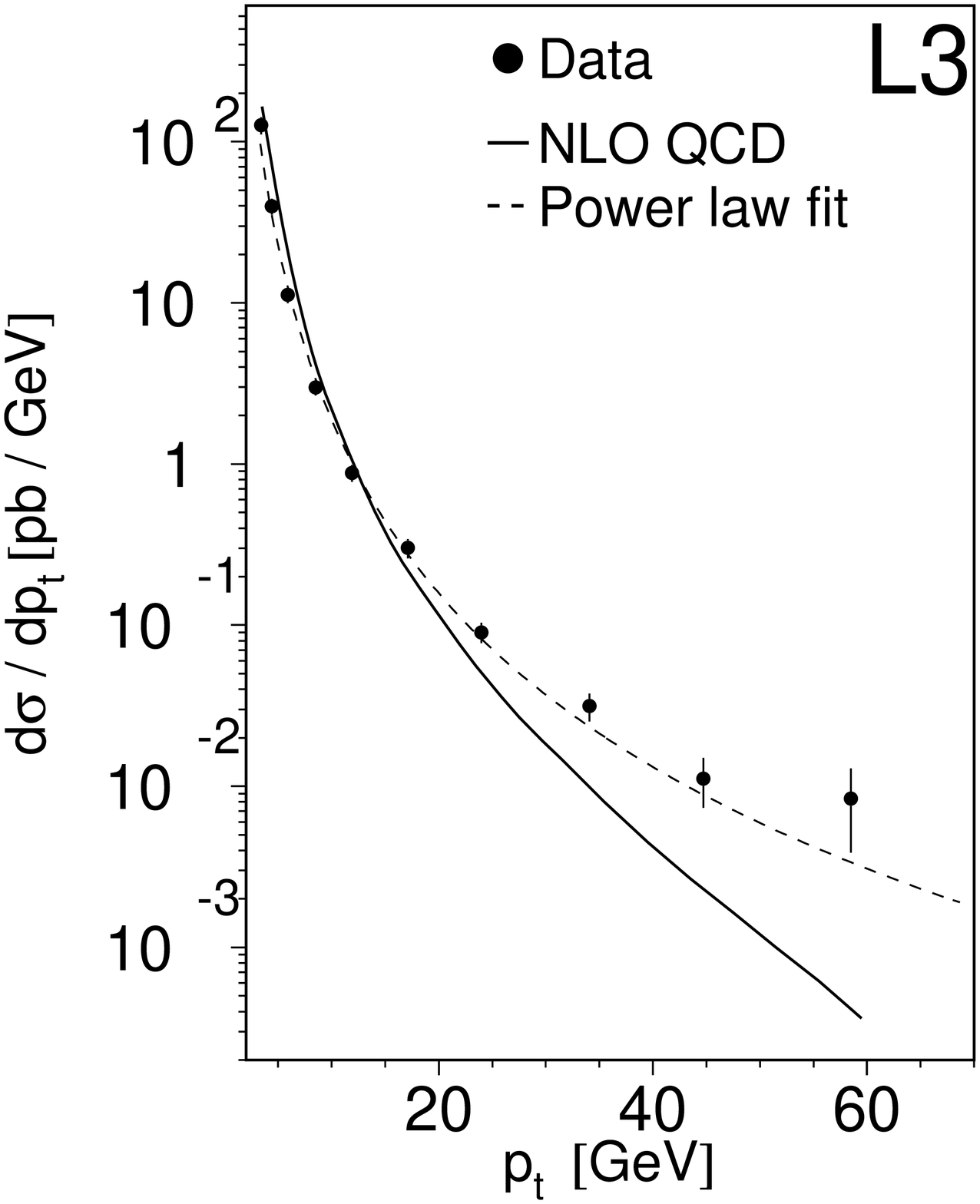}
\end{center}
\caption{The cross section as a function of the transverse momentum for inclusive charged 
hadron (left) and jet (right) production.}
\label{fig02} 
\end{figure}

\section{Inclusive processes}

 The measurement of open charm production in two-photon collisions has been performed
using the lepton tag method by L3~\cite{L3-cc-lep}~\cite{L3-ccbb-lep} and 
the D$^*$ tag method by ALEPH~\cite{ALEPH-cc-D},  L3~\cite{L3-cc-D} and OPAL~\cite{OPAL-cc-D}.
As reported in Figure~\ref{fig01}~(right), good agreement has been found
with perturbative QCD predictions. These results show that only the direct interaction of the
photon is not sufficient to describe the data. The contribution to the cross section
due to the gluonic content of the photon is about 50\%.

 The cross section of open beauty production has been measured by 
L3~\cite{L3-ccbb-lep} and preliminary results have been presented at
by OPAL and DELPHI~\cite{Photon2003}. There is a very good agreement between the three experimental
measurements but, as shown in Figure~\ref{fig01}~(right), an inconsistency of more than a
factor two has been found between data and perturbative QCD predictions. The reason of this 
inconsistency is unknown and a careful investigation of this problem is mandatory.

 The cross section for $\pi^\pm$, $\pi^0$ and ${\rm K}^0_s$ inclusive production in the reaction 
$\gamma\gamma\rightarrow$ hadrons has been measured as a function of the transverse momentum
and the pseudorapidity by L3~\cite{L3-p0k0}~\cite{L3-cha} and OPAL~\cite{OPAL-chak0}. 
Agreement with respect to QCD predictions is found for all particles at $p_t < 4$ GeV
while data are significantly higher than the QCD predictions for $\pi^\pm$ and $\pi^0$
at high $p_t$, as shown in Figure~\ref{fig02}~(left).
 The study of the production of jets in two-photon collisions has been performed
by L3~\cite{L3-jets} and OPAL~\cite{OPAL-jets}. As presented in Figure~\ref{fig02}~(right),
a clear excess of the measured cross section with respect to QCD predictions has been observed
at high $p_t$ values in inclusive jet production by L3.
Several checks have been performed to investigate possible experimental
problems and no effect has been found to be responsible for this discrepancy. A careful
theoretical investigation is therefore mandatory. A natural question rises: 
is there any correlation between these effects at high $p_t$
and the problem of the open beauty production cross section?

\section{Conclusions}

 The LEP collider has given fundamental contributions to two-photon physics.
In particular, the high energy data above the Z pole
allowed the first observation of open beauty 
production and the study of high transverse momentum single particle and jet production.
For these processes significant deviations from QCD
predictions have been observed and no explanation has been found for these
effects at present.

%

\end{document}